\documentclass[%
 reprint,
 amsmath,amssymb,
 aps,
]{revtex4-2}

\usepackage{graphicx}
\usepackage{upgreek}
\usepackage{color}
\usepackage{threeparttable}
\usepackage{dcolumn}
\usepackage{bm}
\usepackage{hyperref}
\hypersetup
{
    colorlinks=true,
    linkcolor=blue,
    filecolor=blue,
    urlcolor=blue,
    citecolor=blue,
}

\begin{document}

\preprint{APS/123-QED}

\title{Buildup dynamics of broadband Q-switched noise-like pulse}

\author{Ji Zhou$^{1,2,3}$} \thanks{These authors contributed equally.}

\author{Yuhang Li$^{1,2,3,*}$}\email{lyhTHU@gmail.com}
\author{Qing Yang$^{3}$} 
\author{Yaoguang Ma$^{3}$}\email{mayaoguang@zju.edu.cn} 
\author{Qiang Liu$^{1,2}$} 
  \affiliation{
    $^{1}$Key Laboratory of Photonic Control Technology (Tsinghua University), Ministry of Education, Beijing 100084, China\\
    $^{2}$State Key Laboratory of Precision Measurement Technology and Instruments, Department of Precision Instrument, Tsinghua University, Beijing 100084, China\\
    $^{3}$State Key Laboratory of Modern Optical Instrumentation, College of Optical Science and Engineering, Zhejiang University,
    Hangzhou 310027, China}
\date{\today}

\begin{abstract}
    
    We investigate the buildup dynamics of broadband Q-switched noise-like pulse (QS-NLP) driven by slow gain dynamics in a microfiber-based passively mode-locked fiber laser. Based on shot-to-shot tracing of the transient optical spectra and qualitatively reproduced numerial simulation, we demonstrate that slow gain dynamics is deeply involved in the onset of such complex temporal and spectral instabilities of QS-NLP. The proposed gain dynamics model could contribute to deeper insight into such nonlinear phenomenon and transient dynamics simulation in ultrafast fiber laser.
                
\end{abstract}

\maketitle

\section{Introduction}
As a typical dissipative system, ultrafast fiber lasers are now increasingly regarded as an ideal platform to study complex dynamics ranging from stationary periodic mode-locking, quasi-periodic pulsating,  exploding and  chaos \cite{Dudley2014Instabilitiesbreathersrogue,Grelu2012Dissipativesolitonsmodea}. Among these numerous nonlinear dynamics, noise-like pulse (NLP) is a typical illustration of partial mode locking which manifests as seemingly regular long pulses (0.1-1 ns, typically) train while actually resembling burst of sub-ps scale chaotic, noise-like inner fluctuations \cite{Horowitz1997Noiselikepulsesbroadband}. Earlier study demonstrated that NLP could circulate with a rather stable envelope at the fundamental roundtrip frequency of the laser cavity, in spite of inner chaotic evolution, while there are also increasing reports recently on Q-switched noise-like pulse (QS-NLP) where Q-switching instabilities are not negligible. In this case, the laser undergoes slow, quasi-periodic energy fluctuations while lasing in a pulsating radiation pattern \cite{Pottiez2019Gaindrivenspectral,Wei201510001400nm,Lecaplain2013Dissipativeroguewave,Zhou2021Broadbandnoisepulse}. When modulated by Q-switching instabilities, QS-NLP usually accumulates energy in a short time, with boosting peak power in the following specific roundtrips and subsequently decay to background noise \cite{Wei201510001400nm,Keller2003Recentdevelopmentscompact}. This temporal intensity dynamics gives rise to nonlinearity enhancement resulting in a broad averaged spectrum that could exceed gain spectral width, while on the other hand contribute to other nonlinear phenomena, e.g. rogue waves (RW)$-$featured with pronounced L-shaped statistics distribution of intensity \cite{Lecaplain2013Dissipativeroguewave,Zhou2021Broadbandnoisepulse}.

Although NLP has been investigated theoretically \cite{Jeong2014formationnoisepulses}, and experimental optical time-stretching \cite{Goda2013DispersiveFouriertransformation,Sun2018Timestretchprobing,Wei201510001400nm} or temporal mapping techniques \cite{Churkin2015Stochasticityperiodicitylocalized} have been successfully in real-time tracing of the NLP dynamic evolution, previous research work focus mainly on the stationary NLP where gain was assumed to saturate instantaneously  with the current energy of pulse \cite{Pottiez2011Adjustablenoiselikepulses,Jeong2014formationnoisepulses,Kwon2017Numericalstudymulti}. This assumption does not agree with the nature of  QS-NLP since Q-switching instabilities related gain evolution usually involves distinct time-scales: pulse duration on a ns scale ($<$0.1 ns here), pulse train on a $\upmu$s scale and slow gain relaxation time on a ms scale (1$-$10 ms) \cite{Giles1991Modelingerbiumdoped,Barnard1994Analyticalmodelrare}. The non-instantaneous slow gain dynamics usually play a pivotal role in several nonlinear dynamics phenomena such as internal oscillations-inhibited stable soliton molecules \cite{Zaviyalov2012Impactslowgain}, gain depletion guided long-range pulses attraction \cite{Weill2016NoisemediatedCasimir}, gain dynamics-dominated interpulse repulsion \cite{Kutz1998Stabilizedpulsespacing},  wavelength-dependent gain competition driven NLP \cite{Pottiez2019Gaindrivenspectral}, as well as QS-NLP here.

Incroporating a slow gain dynamic model, the present work reveals the underlying QS-NLP dynamics in a dispersion and nonlinearity managed fiber ring laser \cite{Zhou2021Broadbandnoisepulse}. Combining real-time dispersive Fourier transform (DFT) measurement of transient spectra and numerical simulation, we experimentally and numerically obtain the temporal and spectral instabilities of QS-NLPs shown similarly to \cite{Wei201510001400nm,Smirnov2017Generationspatiotemporal} and demonstrate that the QS-NLP comes from roundtrip-to-roundtrip dynamic equilibrium between gain and loss. Our work provides  deeper insight into the pulse dynamics of broadband QS-NLP as well as universal gain mechanism in fiber lasers. The proposed model can be used to study active fiber Kerr cavities \cite{Englebert2021Temporalsolitonscoherently,Englebert2021ParametricallydrivenKerr} and on-chip doped microresonator lasers \cite{Kippenberg2006Demonstrationerbiumdoped}.

\section{Experiment}

\begin{figure}[t]
    \includegraphics[width=.45\linewidth]{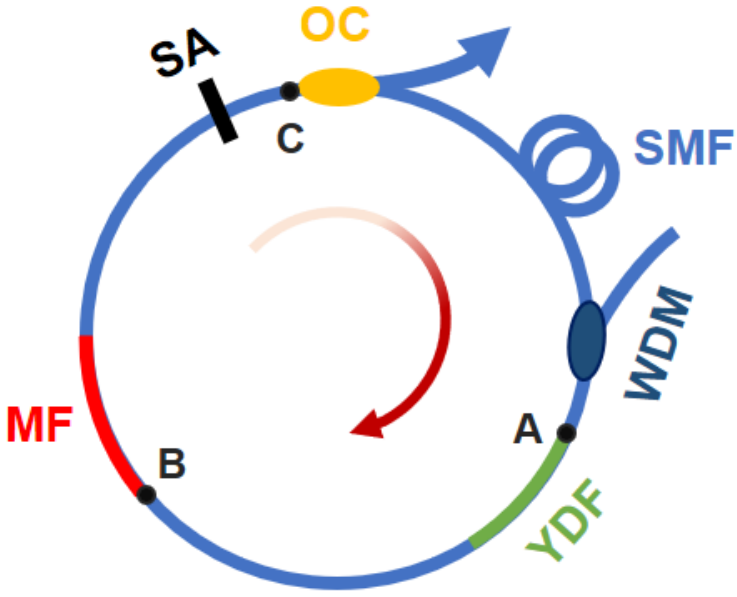}
    \caption{\label{fig:simulation schematic}Schematic configuration of the fiber ring laser in simulation, detailed information see \cite{Zhou2021Broadbandnoisepulse}. MF: microfiber; SMF: Corning$^{\circledR}$ HI1060; YDF: Yb-doped fiber; SA: saturable absorber; OC: output coupler. Intracavity position: A, 0/1.7 m; B, 0.6 m; C, 1.0 m; free space of 0.1 m ignored here.}
\end{figure}

The experiment platform is NPE based fiber ring laser as sketched in Fig. \ref{fig:simulation schematic}. For the DFT measurement, the laser output is temporally stretched by a 1 km fiber with group-velocity dispersion of $\beta_2=20.3 \ \mathrm { ps^2/km}$ at $\sim$1030 nm and subsequently fed into 25 GHz detector (Newfocus Model1414) along with 33 GHz oscilloscope (Tektronix DPO73304DX). As a result, the implemented spectral resolution is $\sim$ 0.9 nm, sufficient for measuring such broadband spectra with 20-dB-bandwidth greater than 100 nm \cite{Goda2013DispersiveFouriertransformation}. 
    
\begin{figure}[b]
    \includegraphics[width=0.49\linewidth]{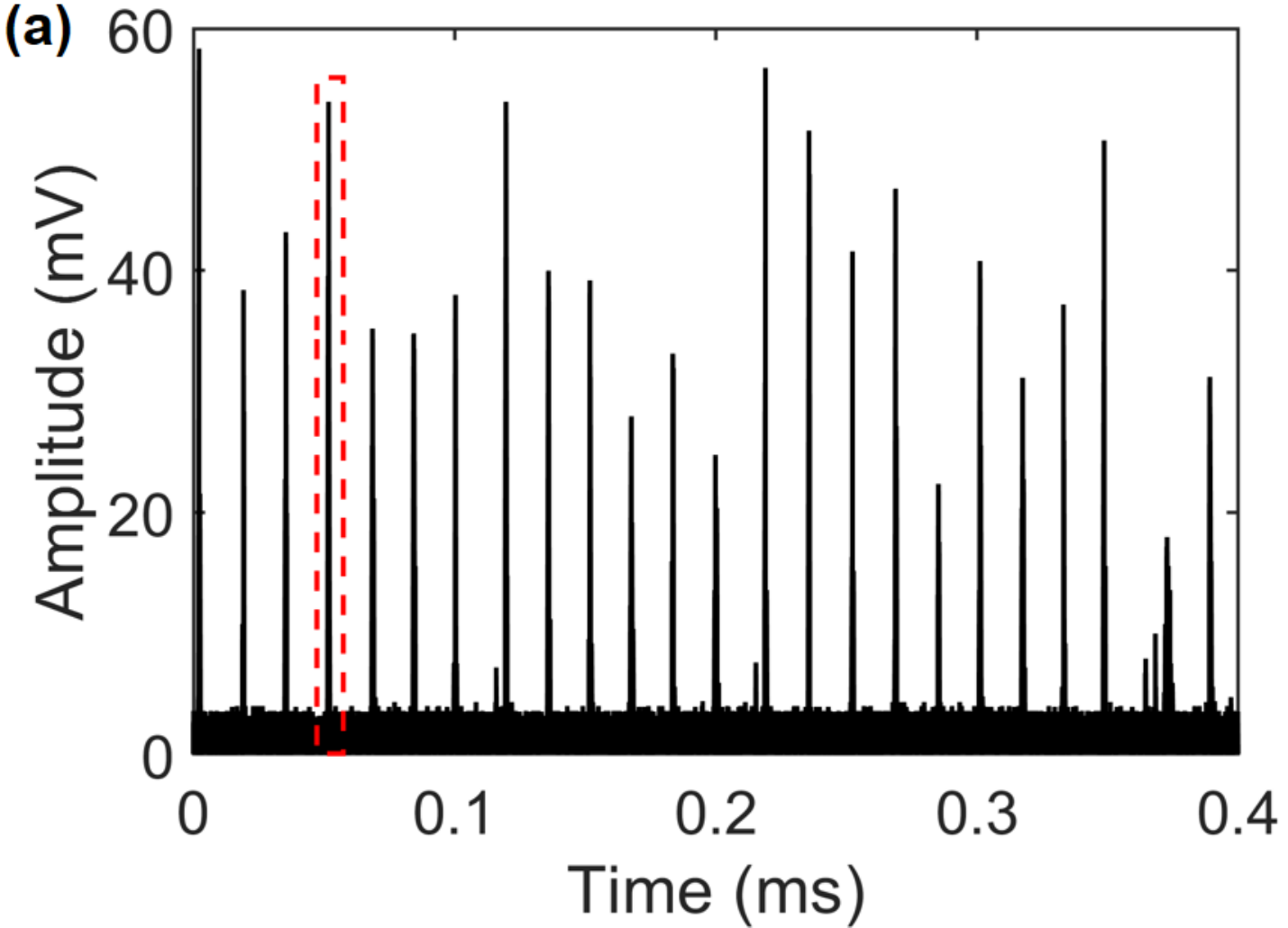}
    \includegraphics[width=0.49\linewidth]{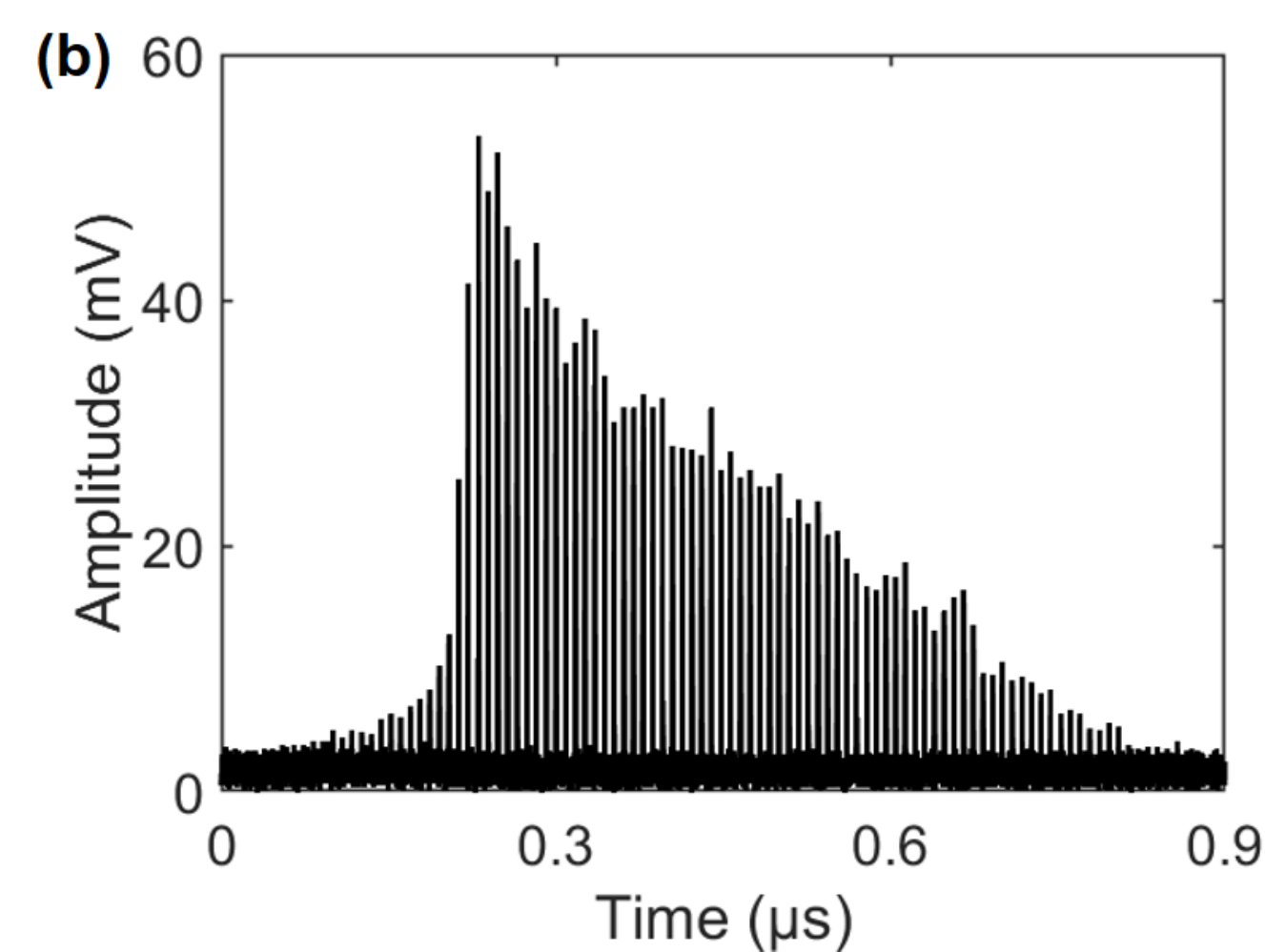}
    \caption{\label{fig:tem DFT result} Temporal measurement results. (a) The temporal train of QS-NLP in ms scale. (b) The zoom-in QS-NLPs intensity dynamics in $\upmu$s scale red boxed in (a), where the pulse power undergoes exponential growth and then decay to negligible level due to strong dissipation. }
\end{figure}

Self-starting QS-NLP operation could be initiated by increasing pump power high than 400 mW with several certain intracavity polarization configuration. We deliberately pump the fiber laser at power of 600 mW to trigger QS-NLP with excess of gain. Fig. \ref{fig:tem DFT result} shows the typical temporal intensity dynamics of QS-NLP in ms and $\upmu$s scale, where cavity-length-determined pulsing (duration of $\sim$8.7 ns) is modulated by Q-switched envelope. At initial few roundtrips, due to stronger bleaching (i.e. positive feedback) effect of SA, the QS-NLP energy grows exponentially from slight relaxation oscillations \cite{Hoenninger1999Qswitchingstability}. The pulse energy grows continuously and thus saturates the gain rapidly. As a result, in the following remaining roundtrips loss has a greater impact than saturated gain and therefore results in to a slowly decaying trail of Q-switched envelope. 

\begin{figure}[t]
    \includegraphics[width=1\linewidth]{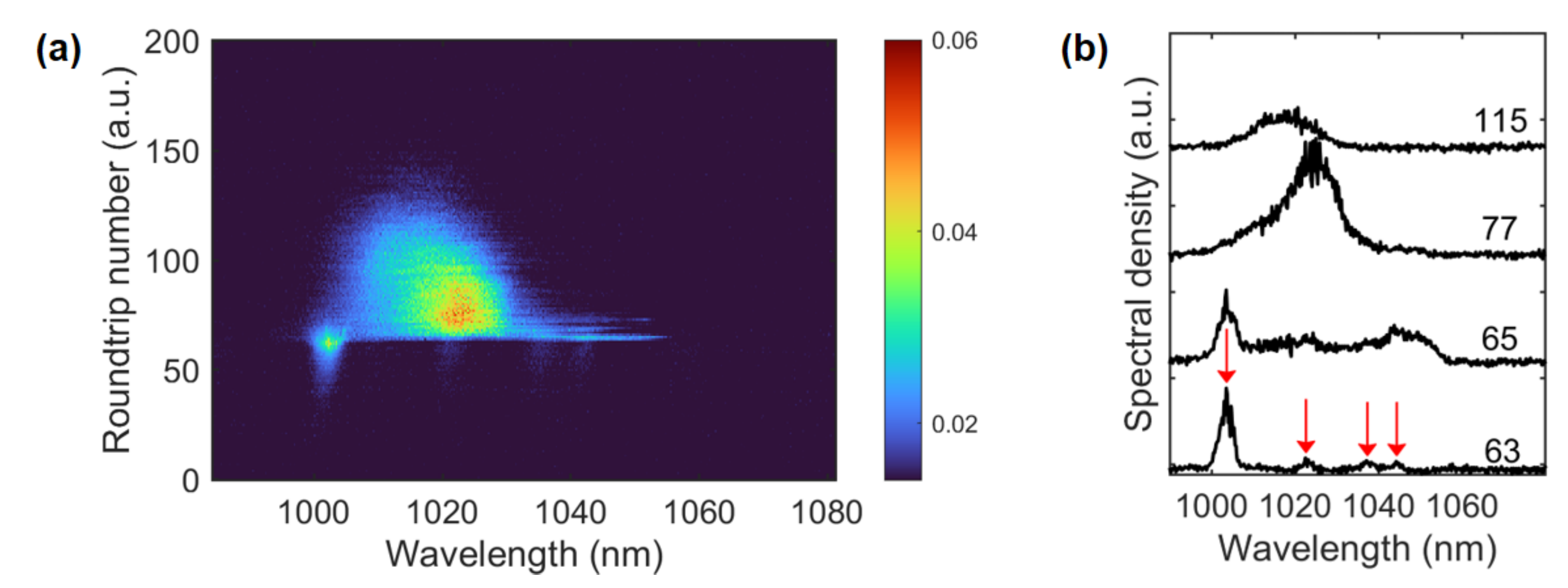}
    \caption{\label{fig:spec DFT result} Typical DFT experiment result of single Q-switched envelope. (a) Single-shot spectra and (b) its sliced spectra at particular roundtrips. The red arrows indicate the initial spectral components related to MI.}
\end{figure}

In the quasi-steady state where sampled averaged spectrum of spectrometer (Agilent 86142B) keeps seemingly fixed, the measured DFT transient spectra is shown in Fig. \ref{fig:spec DFT result} which illustrates typical real-time spectral evolution over hundreds of roundtrips with several obvious features.

At the beginning of such pulsating period, ignited by \textcolor{black}{modulation instability (MI)}, signal pulse grows slightly in several roundtrips. There exist multiple spectral components indicated with red arrows in Fig. \ref{fig:spec DFT result} (b), which is reminiscent of primitive DFT observation in build-up of mode locking \cite{Herink2016Resolvingbuildfemtosecond} and dissipative soliton formation \cite{Peng2018Realtimeobservation}. Moreover, note that MI could be efficiently driven in microfiber of anomalous dispersion \cite{Zhou2021Broadbandnoisepulse}.

Following the above preliminary amplification process, the spectral sideband at $\sim$1030 nm with maximum gain is exponentially amplified in several roundtrips at the expense of pump. Meanwhile,  stemming from the combined high nonlinearity and anomalous dispersion of microfiber  \cite{Zhou2021Broadbandnoisepulse}, the spectra broaden abruptly.

As the pulse energy grows in a short time and meanwhile saturates the gain, the transient broadband spectra in particular of those components beyond gain bandwidth do not survive in the following roundtrips. Furthermore, the spectrum continue narrowing down as long as loss surpass gain. 

In addition to the above DFT measurement, basic characterization of QS-NLP was exhibited in \cite{Zhou2021Broadbandnoisepulse}. But overall, experimental measurement only provides presentational spectral-temporal intensity dynamics, the underlying slow gain dynamics uncovered in earlier solid-state laser experiment \cite{Kaertner1995Controlsolidstate} probably gives rise to the observed pattern formations and we will demonstrate it theoretically.

\section{Simulation modeling}
In this section we give the pulse propagation model in fiber laser incroporating slow gain dynamics. Detailed fiber parameters in particular of microfibers (e.g. when diameter is $\sim$1.2 $\upmu$m, $\gamma=182.2 ~\mathrm{W}^{-1}~\mathrm{km}^{-1}$ and $\beta _2=-156.6~\mathrm{ps}^2/\mathrm{km}$ ) used here refer to \cite{Tong2004Singlemodeguiding,Agrawal2007Applicationsnonlinearfiber} and have been shown in previous experiment \cite{Zhou2021Broadbandnoisepulse}.

\subsection{Gain dynamics modeling}
\label{gain dynamics modeling}

The gain dynamics related rate equations in Yb-doped fiber is modeled like Er-doped amplifiers with two-level character when ignoring the effect of amplified spontaneous emission (ASE) 
\cite{Paschotta1997Ytterbiumdopedfiber,Barnard1994Analyticalmodelrare,Pask1995Ytterbiumdopedsilica}:
\begin{equation}
    \begin{array}{c}
    \left\{\begin{aligned}
    N_{\rm {t}} &=N_{0}(z,t)+N_{1}(z,t) \\
    \frac{\partial N_{1}(z,t)}{\partial t}&=\frac{I_{s}}{h \nu_{s}}\left(N_{0}(z,t) \sigma_{s}^{a}-N_{1}(z,t) \sigma_{s}^{e}\right)+\\&\frac{I_{p}}{h \nu_{p}}\left(N_{0}(z,t) \sigma_{p}^{a}-N_{1}(z,t) \sigma_{p}^{e}\right)-\frac{N_{2}(z,t)}{\tau_{2}}
    \end{aligned}\right.
\end{array}
\label{eq:rate equations}
\end{equation}
where $N_{\rm t}$, $N_0$, $N_1$ are the total population concentration, lower level population and upper level population, respectively. $I_{s}$($I_{p}$), $\nu_{s}$($\nu_{p}$), $\sigma_{s}^{a}$($\sigma_{p}^{a}$), $\sigma_{s}^{e}$($\sigma_{p}^{e}$) are the signal(pump) intensity, frequency, absorption and emission cross-section, $h$ is the Planck constant, $\tau_{2}$ is the upper laser level lifetime.

The gain coefficient $g\left(z, P_{\mathrm{avg}}, \lambda\right)=g_{m}\left(z, P_{\rm {a v g}}\right) g(\lambda)$ when pulse propagates along gain fiber is defined as \cite{Agrawal2007Applicationsnonlinearfiber,Runge2014Allnormaldispersion}:
\begin{equation}
    \left\{\begin{aligned}
    g_{m}\left(z, P_{\rm{a v g}}\right) &=\frac{\partial\left(\ln P_{s}(z)\right)}{\partial z} \\
    &=\varGamma_{s}\left(N_{1}(z) \sigma_{s}^{e}-N_{0}(z) \sigma_{s}^{a}\right) \\
    g(\lambda) &=\mathrm{e}^{-\left(\frac{\lambda-\lambda_{0}}{\Delta \lambda}\right)^{2}}
    \end{aligned}\right.
    \label{gain coefficient amplitude}
\end{equation}\\
in which $g(\lambda)$ is related to Gaussian spectral response near at $\lambda _0 \sim 1030$ nm with gain bandwidth $\Delta\lambda$, $\varGamma_s$ is the signal overlap integral \cite{Giles1991Modelingerbiumdoped,Barnard1994Analyticalmodelrare} and $P_{\rm {a v g}}$ corresponds to the average power along the gain fiber.

When assuming a constant, uniform pump along the gain fiber (although impractical, as shown in \cite{Pottiez2019Gaindrivenspectral}, it does not obscure the physical processes to explain),  we can derive the fast gain dynamics equation similar to \cite{Pottiez2019Gaindrivenspectral} after combining Eq. (\ref{eq:rate equations}) and Eq. (\ref{gain coefficient amplitude}):
\begin{equation}
    \frac{\partial g_{m}}{\partial t}=-\left(\frac{1}{\tau_{\mathrm{e}}}+\frac{ P_{s}}{P_{s,\mathrm{sat}} \cdot \tau_{\mathrm{e}}}\right) g_{m}+\Lambda
    \label{fast gain dynamics Eq}
\end{equation}
where $\tau _{\mathrm{e}} =\tau _2/\left(1+P_p/P_T\right)$ is the effective upper laser level lifetime with pump transparency power $P_{T}=h\nu _p A_{\mathrm{eff},p}/\left[ \left( \sigma _{p}^{a}+\sigma _{p}^{e} \right) \tau _2 \right]$ \cite{Barnard1994Analyticalmodelrare} and pump power $P_p=I_p A_{\mathrm{eff},p}=h\nu _p\phi _pA_{\mathrm{eff},p}$; $P_{s,\mathrm{sat}}=h\nu _sA_{\mathrm{eff} ,s}/\left[ \left( \sigma _{s}^{a}+\sigma _{s}^{e} \right) \tau _{\mathrm{e}} \right] $ is the signal saturation power \cite{Pottiez2019Gaindrivenspectral}; the constant $\Lambda=-\varGamma _sN_{\mathrm{t}}\sigma _{s}^{a}/\tau _2-\varGamma _s\phi _pN_{\mathrm{t}}\left( \sigma _{s}^{a}\sigma _{p}^{e}-\sigma _{s}^{e}\sigma _{p}^{a} \right)$  accounts for the influence of pump in excess, $\Lambda \cdot \tau_{\mathrm{e}}$ is equivalent to the well-known small-signal gain coefficient \cite{Agrawal2007Applicationsnonlinearfiber}; and $P_{s}$ is the signal power.

It is worth noting that similar gain dynamics formula  proposed in \cite{Agrawal2007Applicationsnonlinearfiber} with realistic upper level lifetime $\tau_2$ is replaced by $\tau_{\mathrm{e}}$ in Eq. (\ref{fast gain dynamics Eq}) and there is usually two orders of magnitude difference between them. This pump induced transient response within effective lifetime $\tau_{\mathrm{e}}$ can be traced back primitive research of EDFA in WDM systems \cite{Sun1999Analyticalformulatransient,Desurvire1989Analysistransientgain}.

In the mean-field model, the fast gain dynamics that mainly affects the inner structure of pulse wavepacket \cite{Zaviyalov2012Impactslowgain} is neglected and we only consider the roundtrip-to-roundtrip slow  gain dynamics in the unit of round-trip time $T_R$. Based on multi-scale perturbation approach  \cite{Haboucha2008Analysissolitonpattern}, we obtain the similar slow gain dynamics equation proposed in \cite{Englebert2021Temporalsolitonscoherently,Englebert2021ParametricallydrivenKerr} expressed as:
\begin{equation}
    T_{\mathrm R}\frac{\partial g_m}{\partial T}=-\left( \frac{T_{\mathrm R}}{\tau _{\mathrm{e}}}+\frac{\int_0^{T_{\mathrm R}}{|}A(T,t)|^2\mathrm{d}t}{P_{s,\mathrm{sat}}\cdot \tau _{\mathrm e}} \right) g_m+T_{\mathrm R} \cdot \Lambda 
    \label{mean-field model Eq}
\end{equation}
in which $T=n \cdot T_{\mathrm R}$ is the slow-time variable and $n$ is an integer. 

Furthermore, in the slow gain form Eq. (\ref{mean-field model Eq}) could be written as:
\begin{equation}
    \frac{\partial g_{m}}{\partial T}=-\left(\frac{1}{\tau_{e}}+\frac{\left\langle P_{s}\right\rangle}{P_{s, s a t} \cdot \tau_{e}}\right) g_{m}+\Lambda
    \label{slow g}
\end{equation}
where $\left\langle P_{s}\right\rangle=\frac{1}{T_{\mathrm R}} \int_{0}^{T_{\mathrm R}}|A(T, t)|^{2} \mathrm{d} t$ is intracavity signal average power. Eq. (\ref{slow g}) is the ultimate equation describing the roundtrip-to-roundtrip slow gain dynamics for simulation.

Besides, the utilized parameters of YDF \cite{Barnard1994Analyticalmodelrare,Pask1995Ytterbiumdopedsilica,Paschotta1997Ytterbiumdopedfiber} for modeling are listed in Table \ref{tab:parameters for rate equation}.
\begin{table}[hbpt]
    \centering
    \caption{Simulation parameters of YDF}
    \label{tab:parameters for rate equation}
    \begin{threeparttable}
    \begin{tabular}{ccc}
    \hline
    Parameter                        & Value                                  \\ \hline
    length, $l$                       & 0.2 m                                  \\
    dopant radius, $b$               & 2.0\tnote{*} $\upmu$m \\
    pump power, $P_p$                & 0.6 W                                  \\
    upper level lifetime, $\tau_{2}$ & 770 $\upmu$s                           \\
    $\mathrm {Y b^{3+}}$ concentration, $N_{\rm{t}}$                             & $1.2 \times 10^{26}$\tnote{*} $\mathrm{m}^{-3}$ \\
    mode field radius, $w_s$/$w_p$   & 2.2/2 $\upmu$m                         \\
    overlap integral, $\varGamma_{s}$/$\varGamma_{p}$           & 0.56/0.63                                                        \\
    absorption cross section, $\sigma_{p}^{a}$/$\sigma_{s}^{a}$ & $2.5/0.05 \times 10^{-24}$ $\mathrm{m}^{2}$                      \\
     emission cross section, $\sigma_{p}^{e}$/$\sigma_{s}^{e}$   & $2.5/0.6 \times 10^{-24}$ $\mathrm{m}^{2}$                       \\ \hline
    \end{tabular}
    \begin{tablenotes}
        \footnotesize
        \item[*] Undisclosed values.
      \end{tablenotes}
  \end{threeparttable}
\end{table}

\subsection{Pulse propagation modeling}

The schematic configuration of the fiber laser is shown in Fig.\ref{fig:simulation schematic}, which consist of a section of 0.2 m microfiber with diameter of 1.2 $\upmu$m, a piece of 0.2 m Yb-doped fiber (LIEKKI$^{\circledR}$ Yb1200-4/125) and remaining of 1.3-m-long SMF (HI1060).

We model the fiber laser with generalized nonlinear Schrodinger equation \cite{Runge2014Allnormaldispersion}, i.e.,
\begin{equation}
\begin{aligned}
    \label{eq: NLSE}
    &\frac{\partial A}{\partial z}-\sum_{k=2}^{4} \frac{i^{k+1}}{k !} \beta_{k} \frac{\partial^{k} A}{\partial t^{k}}-\frac{g\left(z, P_{\text {avg }}, \omega\right)}{2} A \\
    &=i \gamma\left(1+\frac{i}{\omega_{0}} \frac{\partial}{\partial t}\right)\left[A(z, t) \int_{-\infty}^{\infty} R\left(t^{\prime}\right)\left|A\left(z, t-t^{\prime}\right)\right|^{2} \mathrm{d} t^{\prime}\right]
\end{aligned}
\end{equation}
where terms on the right hand side responsible for stimulated Raman scattering (SRS) and SPM are included; $g\left( z,P_{\mathrm{avg}},\mathrm{\omega} \right) $ is the gain coefficient (otherwise set as zero for passive fiber) described by Eq. (\ref{gain coefficient amplitude}) and Eq. (\ref{slow g}).

For convenience, in simulation we only foucus on the slow gain dynamics at signal wavelength ($\sim$1030 nm) and ignore other spectral components even though whose dynamic evolution exist as shown in Fig. \ref{fig:spec DFT result}.
\begin{figure}[t]
    \includegraphics[width=0.7\linewidth]{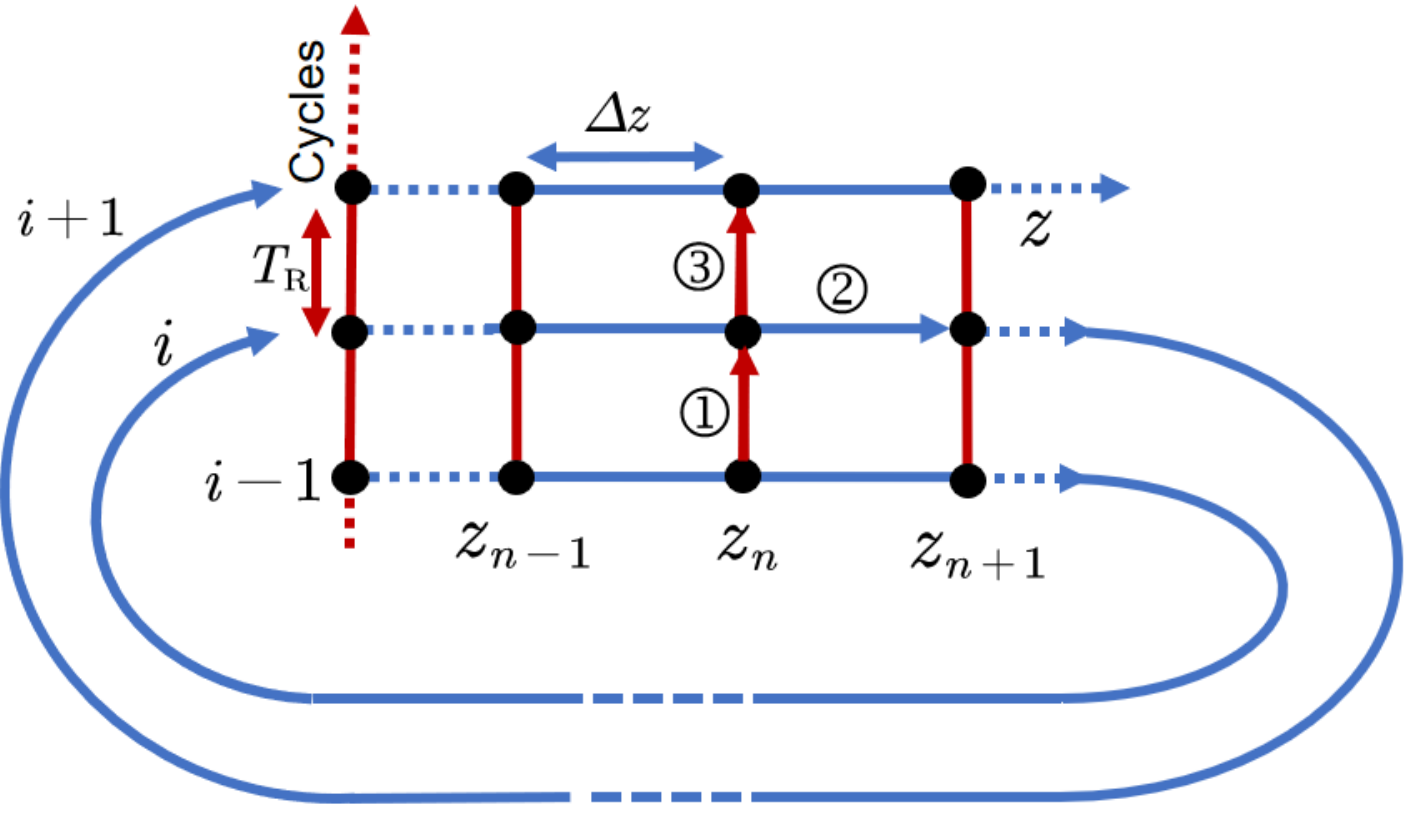}
    \caption{\label{numerical model} Schematic of the numerical calculation procedure of pulse propagation in YDF. Red lines correspond to gain evolution over cycles in the unit of $T_{\mathrm{R}}$ while the blue lines associated with pulse propagation in each cycle. Circled numbers indicate the calulation steps at each node.}
\end{figure}

Numerical treatment of single pulse propagation in YDF is similar to \cite{Pottiez2019Gaindrivenspectral} and shown in Fig. \ref{numerical model}. Before considering the pulse profile $A(T_{i}, z_n)$ at the point $z_{n}$ in $i$-th cycle, the gain $g(T_{i}, z_n)$ should be evaluated by integrating Eq. (\ref{slow g}) over $T_{\mathrm R}$ in advance, as expressed in Eq. (\ref{eq: integral equations}). Afterwards such gain coefficient allows determining the pulse profile at $z_{n+1}$ with Eq. (\ref{eq: NLSE}). Besides, the pulse profile $A(T_{i}, z_{n})$ determines the local averaged power $\left\langle P_{s}(z_{n})\right\rangle$, which in turn influence the $g(T_{i+1}, z_{n})$ in next cycle. We can repeat the procedure to simulate the buildup dynamics of QS-NLP in such fiber laser.
\begin{equation}
    \left\{\begin{array}{l}
    \label{eq: integral equations}
    g\left(T_{i}, z_{n}\right)=g\left(T_{i-1}, z_{n}\right)+\int_{T_{i-1}}^{T_{i}} \frac{\partial g\left(T_{i-1}, z_{n}\right)}{\partial T} \mathrm{~d} T \\
    A\left(T_{i}, z_{n+1}\right)=A\left(T_{i}, z_{n}\right)+\int_{z_{n}}^{z_{n+1}} \frac{\partial A\left(T_{i}, z_{n}, g\left(T_{i}, z_{n}\right)\right)}{\partial z} \mathrm{~d} z
\end{array}\right.
\end{equation}

Prior to simulation, we set the initial gain to the small-signal value at all position along the YDF. An sech-shape small signal (peak power of 1 $\upmu$W) plusing Gaussian background noise is fed into fiber laser initially to accelerate the convergence of calculation.

Finally, typical parameters value of intracavity elements are listed as follow \cite{Senel2018Tailoreddesignmode,Zhou2021Broadbandnoisepulse}:

(i) YDF: $\gamma =11.5 ~\mathrm{W}^{-1}~\mathrm{km}^{-1}$; $\beta _2=26.2~\mathrm{ps}^2/\mathrm{km}$, $\beta _3 = -0.0134~\mathrm{ps}^3/\mathrm{km}$.

(ii) SMF: $\gamma =5.9 ~\mathrm{W}^{-1}~\mathrm{km}^{-1}$; $\beta _2=24.8~\mathrm{ps}^2/\mathrm{km}$, $\beta _3 = -0.0233~\mathrm{ps}^3/\mathrm{km}$.

(iii) MF: $\gamma=182.2 ~\mathrm{W}^{-1}~\mathrm{km}^{-1}$; $\beta _2=-156.6~\mathrm{ps}^2/\mathrm{km}$, $\beta_3=0.152 ~\mathrm{ps}^3/\mathrm{km}$, $\beta_4 =1.55 \times 10^{-4}~\mathrm{ps}^4/\mathrm{km}$.

(iv) SA: $T=1-l_0/\left[ 1+P_s (t)/P_{\mathrm{sat}} \right]$ with $l_0 = 0.84$ and $P_{\mathrm{sat}} = 90$ W. 

\section{results and discussion}
Experimentally, as mentioned before that the QS-NLP could easily self-start at certain intracavity polarization, i.e. specific SA transmission function $T$. It is well-known that SA always provides positive feedback in the self-starting mode-locking process where the main pulses strat up from initial noise fluctuation formed by mode-beating \cite{Nelson1997Ultrashortpulsefiber}, while NLP self-starting process usually involves subsequently futher increment in the pulse power that instead lead to negative feedback of SA \cite{Jeong2014formationnoisepulses}. In the condition of high linear intracavity loss, only at these locations where previous main pulses disturbed can the dispersive wave, collapsed soliton components or background noise generate, undergo amplification and further circulate as bunch temporally \cite{Tang2005Solitoncollapsebunched,Zhao2007Noisepulsegain}. As a result, such low coherence components bunched NLP presents typical double-scale (fs-ps) temporal autocorrelation profile \cite{Zhou2021Broadbandnoisepulse} and apparent spectral roundtrip-to-roundtrip fluctuations\cite{Lecaplain2014Roguewavesnoiselike,Runge2013Coherenceshotshot}.

\begin{figure}[b]
    \includegraphics[width=1\linewidth]{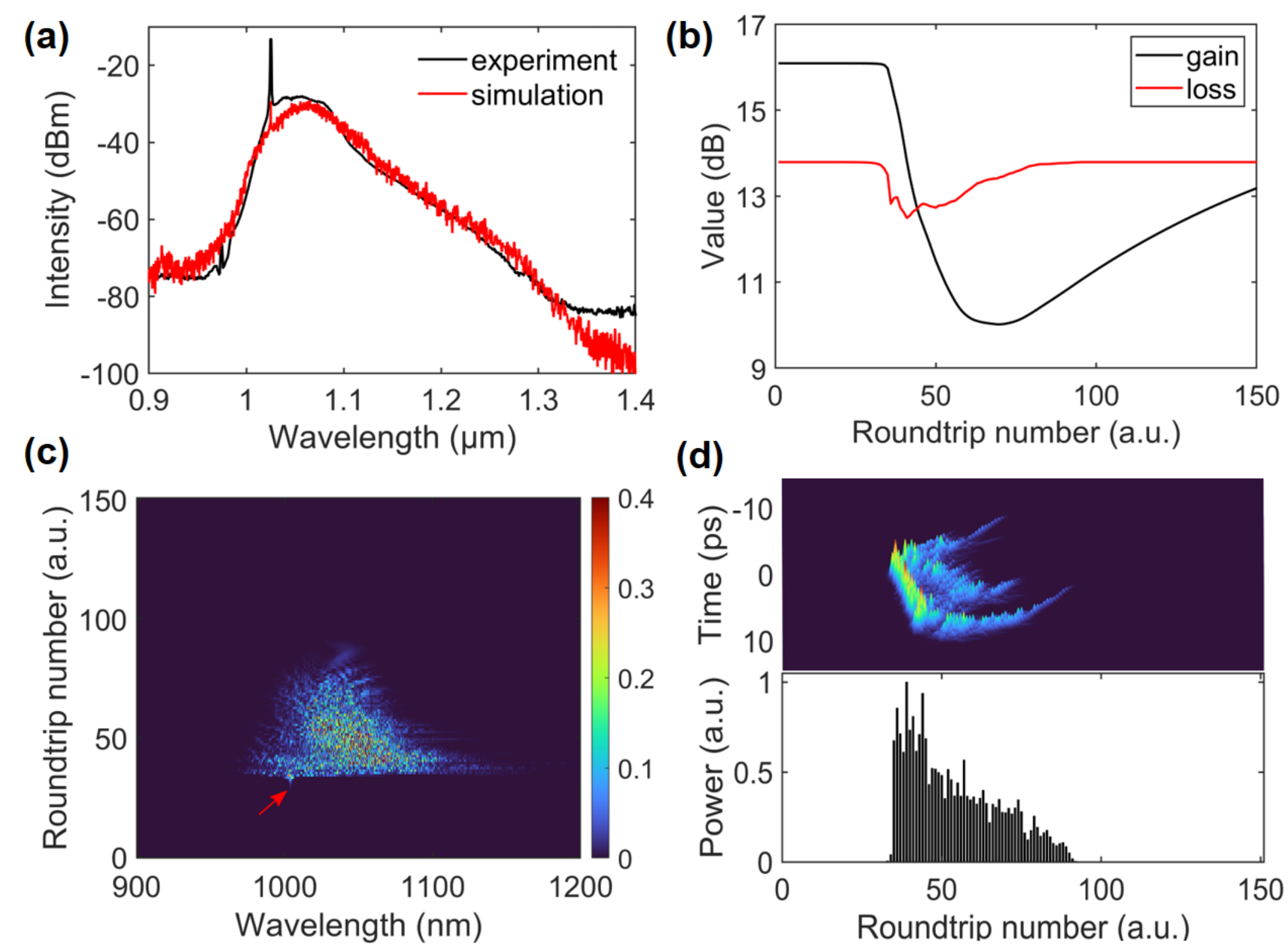}
    \caption{\label{fig:simulation figs}Simulation of single QS-NLP dynamics over 150 roundtrips. (a) Averaged simulation spectrum and experimental result. (b) Gain/loss level evolution over round trips, there  come into being QS-NLPs when gain surpass loss \cite{Kaertner1995Controlsolidstate}. (c) Simulated spatio-temporal spectral dynamics in linear scale, the red arrow indicates the initial small signal. (d)Upper panel: temporal intensity dynamics over roundtrips;
    lower panel: pulse peak value evolution over roundtrips.}
\end{figure}

We emphasize that the laser always radiates quasi-periodically (intuitive display refers to Fig. 3 (a) in \cite{Zhou2021Broadbandnoisepulse}). In simulation, for simplicity we only focus on laser dynamics within single Q-switched envelope. 
Fig. \ref{fig:simulation figs} shows the typical numerical results in hundreds of roundtrips. We could observe that the simulated spectral profile fits well to the experimental one in Fig. \ref{fig:simulation figs} (a). 

According to Fig. \ref{fig:simulation figs} (b), we can directly observe the underlying dynamic evolution of both gain and loss(counts the influence of SA and OC) within Q-switched envelope. Considering together with Fig. \ref{fig:simulation figs} (d), we can find that at the amplification stage the saturatable gain and loss decrease synchronously along with rapid growth of pulse energy. After that loss prevails over gain, hinges with pulse power and thus leads to a slowly decaying trail of Q-switched envelope.

The spectral dynamic evolution is shown in Fig. \ref{fig:simulation figs} (c). We can find the spectrum could broaden abruptly and then narrows down, which is qualitatively similar to the experimental observation in Fig. \ref{fig:spec DFT result} (c) although ignoring initial multiple spectral components.

Lastly, as we only consider the artifical SA in a quite simple generalized formula, the simulation is unable to fully reproduce the experimental observation and insufficient for accurate exploration of other unconcerned experimental valuable parameters such as Q-switched envelope period and Q-switching stability limits \cite{Hoenninger1999Qswitchingstability,Cheng2016Theoreticalexperimentalanalysis}. Further detailed experimental parameters access by either intracavity precise polarization management \cite{Pu2019Intelligentprogrammablemode} or changing the laser structure \cite{Runge2014Allnormaldispersion} will be helpful for deeper exploring of the relevant underlying mechanism.  

\section{Conclusion}

In summary, based on a simple gain dynamics model we demonstrate that, in  our highly dissipative fiber laser, the Q-switched characteristic of NLP is attributed to the combined impact of gain dynamics (saturable gain depletes rapidly whereas recovers slowly) and SA effect. 

The proposed convenient numerical model here we believe that not only provides essential insight into the operational dynamics of experimentally realized architectures like RWs \cite{Zhou2021Broadbandnoisepulse,Zaviyalov2012Roguewavesmode,SotoCrespo2011Dissipativeroguewaves} and other nonlinear phenomena involving Q-switching instabilities \cite{Wang2018simultaneousgenerationsoliton, Qiao2016GenerationQswitched}, but also allows for simple and fast exploration of various parameter regimes so as to discover new laser designs with improved performance, e.g. high peak power broadband light source \cite{Chernikov1997SupercontinuumselfQ}.
\vspace{0.2cm}

\newpage
\begin{acknowledgments}
    We thank Wenbin He for inspired discussions. This work was supported by the National Key Research and Development Program of China (Grants No. 2020YFC2200403), National Natural Science Foundation of China (Grants No. 62175122 and Grants No. 61905213).
\end{acknowledgments}

\appendix
\section{Gain dynamics over consecutive Q-switched envelopes}
\begin{figure}[h]
    \includegraphics[width=1\linewidth]{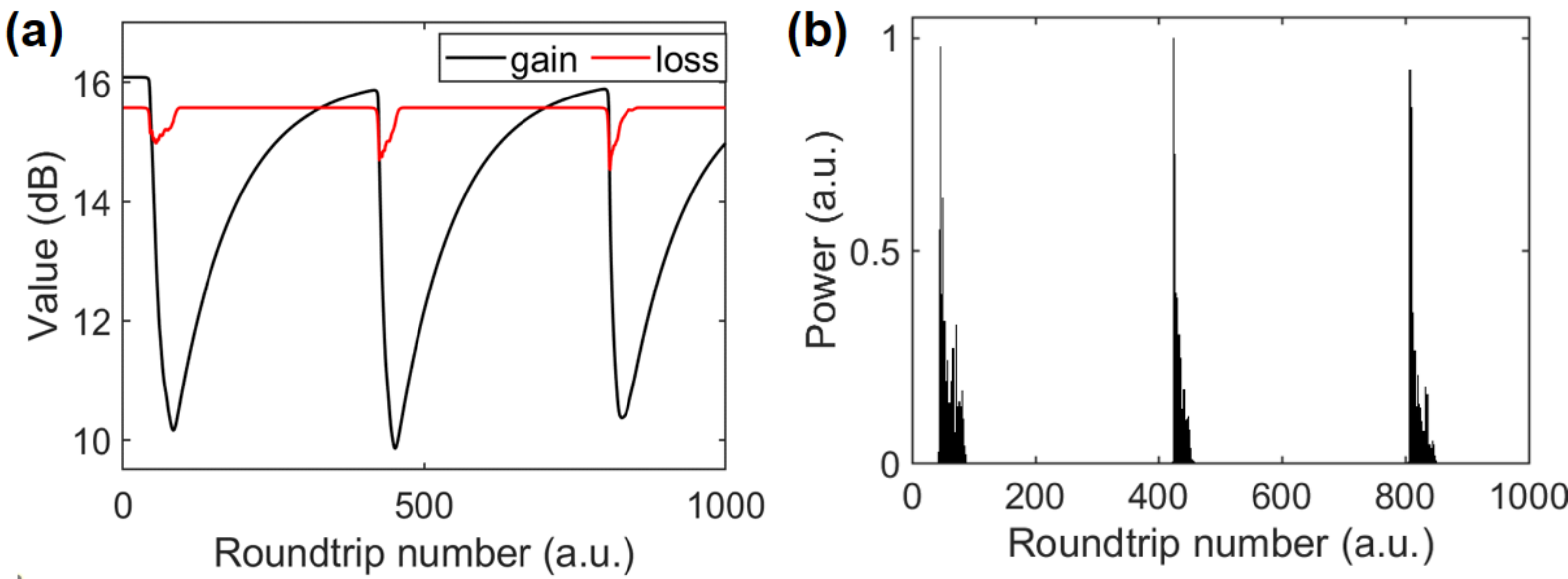}
    \caption{\label{long gain dynamics} (a) Gain/loss evolution over sequences of Q-switched envelopes. (b) Temporal QS-NLP intensity dynamics.}
\end{figure}

In order to provide more intuitive insight into the slow gain dynamics over much longer roundtrips than shown in Fig. \ref{fig:simulation figs}. Here we also calculate the envelope-to-envelope gain dynamics shown in Fig. \ref{long gain dynamics}. We can see the pronounced dynamic evolution of slow gain as well as fast saturation absorption loss. 

\nocite{*}
\bibliography{apssamp1}
\end{document}